\documentclass[%
 aip,
 jcp,%
 amsmath,amssymb,
 reprint,%
]{revtex4-1}
\usepackage{graphicx}
\usepackage{amsmath}
\usepackage{amssymb}
\usepackage{dcolumn}
\usepackage{bm}

\newcommand{\dd}{\textrm{d}}

\usepackage{color}
\begin{document}

\title{Quantum Monte Carlo with Variable Spins}

\author{Cody A. Melton$^{1}$, M. Chandler Bennett$^{1}$, and Lubos Mitas$^{}$}
\affiliation{
1) Department of Physics, North Carolina State University, Raleigh, North Carolina 27695-8202, USA\\
}

\date{\today}

\begin{abstract}
We investigate the inclusion of variable spins in electronic structure quantum Monte Carlo, with a focus on diffusion Monte Carlo with Hamiltonians that include spin-orbit interactions. Following our previous introduction of fixed-phase spin-orbit diffusion Monte Carlo (FPSODMC), we thoroughly discuss the details of the method and elaborate upon its technicalities. We present a proof for an upper-bound property for complex nonlocal operators, which allows for the implementation of T-moves to ensure the variational property. We discuss the time step biases associated with our particular choice of spin representation. Applications of the method are also presented for atomic and molecular systems. We calculate the binding energies and geometry of the PbH and Sn$_2$ molecules, as well as the electron affinities of the 6$p$ row elements in close agreement with experiments. 
\end{abstract}

\maketitle

\section{Introduction}
Quantum Monte Carlo (QMC) methods have become an important tool in understanding the electronic structure for a multitude of systems including atoms, molecules, clusters and solids. In particular,  variational Monte Carlo (VMC) and diffusion Monte Carlo (DMC) have had great success in capturing the many-body correlation effects that influence material properties such as binding and cohesive energies, excitations, phase transitions, etc; these properties are calculated to high accuracy and are in excellent agreement with experiment \cite{qmcrev,KolorencMitasReview}. DMC is a projector method
that applies the operator $\exp(-\tau H)$, where $H$ is the system Hamiltonian, to an appropriate trial  or variational wave function. In the imaginary time limit $\tau \rightarrow \infty$, the ground state for a given symmetry is obtained while excited state contributions to the trial wave function are exponentially damped. Plagued by the fermion sign problem, DMC applications to electronic systems often invoke the fixed-node (FNDMC) approximation \cite{anderson75,anderson76,reynolds82} which fixes the nodal structure of the solution to that of an appropriate trial wave function, which is typically obtained from anti-symmetric combinations of (post-)Hartree-Fock (HF) or Density Functional Theory (DFT) one-particle orbitals. In the case where the trial nodes are exact, the exact ground state energy of the Hamiltonian is obtained. Approximate nodes yield variational estimates of the energy that in many cases proved to be remarkably accurate  even for single-reference trial functions based on DFT or HF orbitals.

Despite its successes, FNDMC actually solves only the spatial part of the eigenstate problem since in typical calculations the electronic spins are treated statically rather than as true quantum variables. This is perfectly adequate in systems where the Hamiltonian does not include spin, and then the nontrivial part of the solution is reduced to spatial dependences only. The particular spin configuration of individual electrons is conserved and therefore is imposed as a symmetry of the system, e.g., a singlet or triplet state. However, many interesting systems exhibit interactions between the spin and spatial degrees of freedom such as the spin-orbit interaction. For nuclear systems, the quantum nature of spins have been realized in variational\cite{carlson} and auxillary field Monte Carlo\cite{sarsa,pederiva,gandolfi} methods. In condensed matter applications, a DMC method was implemented for the 2D homogeneous electron gas with Rashba interaction \cite{ambrosettidmc} as well as its modification to the VMC method applied to atoms\cite{ambrosettivmc}. In these two approaches, the spinor states are stochastically sampled as opposed to sampling the particle coordinate space
that underlies the DMC methods. Recently, we introduced a DMC method which keeps the trial spinors intact during the imaginary time evolution \cite{melton}. This method is particularly useful in that it has the zero-variance property, namely that for arbitrary configurations (spin and spatial coordinates) the bias in energy is proportional to the square of the trial wave function error. Since spin-orbit is  nonlocal in the particle coordinate space, the method deals with spin-orbit terms in a manner similar to nonlocal pseudopotentials 
\cite{mitas91,casula06} and,
as we show below, many of the developed techniques then carry over.  

In this paper, we elaborate and expand upon the details and technical issues of the method introduced previously \cite{melton}. In \S \ref{section:FPDMC}, we give a short discussion of the fixed-phase method in a configuration space without the spin degrees of freedom. In later sections, we generalize the  fixed-phase algorithm to incorporate varying spins and spin-dependent Hamiltonians. In \S \ref{section:spinorbit}, we introduce one particular choice of spin-dependent Hamiltonian, namely the spin-orbit interaction via a pseudopotential. We note that any spin-dependent Hamiltonian could be included, however we choose to focus on the spin-orbit interaction.  In \S \ref{section:spin}, we discuss the inclusion of the spin variables and our choice for the spin representation. In \S \ref{section:errors}, we discuss corresponding timestep dependences in evolutions of both spatial and spin degrees of freedom. We present applications of the method to several atoms and molecules in \S \ref{section:applications}. We conclude in \S \ref{section:conclusions}. 

\section{Fixed-Phase Diffusion Monte Carlo}\label{section:FPDMC}
In order to include spin-dependent Hamiltonians within DMC, we must work with inherently complex wave functions. FNDMC nominally treats real-valued wave functions, so we must resort to a generalization 
of the method. Before dealing with spin terms, we present a short review of  
the fixed-phase method (FPDMC) \cite{ortiz} and its relation to the fixed-node flavor of the DMC. 

For an $N$-electron system, we work in a configuration space $\mathbf{R} = (\mathbf{r}_1, \mathbf{r}_2, \ldots, \mathbf{r}_N) \in \mathbb{R}^{dN}$, where $d$ is dimensionality and here we assume $d=3$. For now, we consider particle spin to be a label rather than a variable. We assume the Born-Oppenheimer approximation, so that we have the Hamiltonian of the form $H=-(1/2)\nabla^2 + V(\mathbf{R})$, where $\nabla = (\nabla_1,\nabla_2, \ldots, \nabla_N)$ and $V$ denotes the electron-ion and electron-electron Coulomb interactions. Since the wave function $\Psi(\mathbf{R},\tau)$ is complex, we write $\Psi(\mathbf{R},\tau) = \rho(\mathbf{R},\tau)e^{i \Phi(\mathbf{R},\tau)}$ and substitute into the imaginary-time Schr\"{o}dinger equation. This yields two coupled differential equations for the amplitude, $\rho(\mathbf{R},\tau)$, and phase, $\Phi(\mathbf{R},\tau)$, as 
\begin{eqnarray}
    \label{rho-eqn} -\frac{\partial \rho(\mathbf{R},\tau)}{\partial \tau} &=& \left[ T_{kin} + V(\mathbf{R}) + \frac{1}{2} \left| \nabla \Phi(\mathbf{R},\tau)\right|^2\right] \rho(\mathbf{R},\tau) \nonumber \\ \\
    -\frac{\partial \Phi(\mathbf{R},\tau)}{\partial \tau} &=& \left[ T_{kin}+ \frac{\nabla \rho(\mathbf{R},\tau) \cdot \nabla}{\rho(\mathbf{R},\tau)} \right] \Phi(\mathbf{R},\tau)
    \label{phase-eqn}
\end{eqnarray}
where we abbreviate $T_{kin}=-(1/2)\nabla^2$.
In order to obtain an approximate solution, we invoke the fixed phase approximation by taking $\partial_\tau \Phi(\mathbf{R},\tau) = 0$ with $\Phi(\mathbf{R},0) = \Phi_T(\mathbf{R})$, where $\Phi_T(\mathbf{R})$ is the phase of a trial or guiding wave function. Writing the trial wave function as $ \Psi_T (\mathbf{R})= \alpha(\mathbf{R})+ i \beta(\mathbf{R})$, we obtain one expression for the fixed trial amplitude and phase 
\begin{eqnarray}
    \label{amplitude}
    \rho_T(\mathbf{R}) &=& \sqrt{\alpha^2(\mathbf{R})+\beta^2(\mathbf{R})} \\
    \label{phase}
    \Phi_T(\mathbf{R}) &=& \tan^{-1}\frac{\beta(\mathbf{R})}{\alpha(\mathbf{R})}
\end{eqnarray}
so that $-\pi/2\le \Phi_T(\mathbf{R}) \le \pi/2$. Since the overall constant phase is irrelevant we can 
alternatively define
\begin{equation}
\Phi_T(\mathbf{R}) =\cot^{-1}\frac{\beta(\mathbf{R})}
{\alpha(\mathbf{R})}
\end{equation}
so that we get  $0\le \Phi_T(\mathbf{R}) \le \pi$.
 The stationary phase condition makes the equation (\ref{phase-eqn}) moot, while the equation for the non-negative amplitude determines the energy eigenvalue. 
 
 \subsection{Fixed-phase upper bound property}
The fixed-phase approximation is variational since the repulsive potential $1/2\left| \nabla \Phi_T(\mathbf{R}) \right|^2$ can only raise the energy for an approximate phase \cite{ortiz}. 
 This is easy to see from the energy expectation with $\rho\exp(i\Phi_T)$ that must be an upper 
bound to the exact energy for
an arbitrary symmetric $\rho\geq 0$. 

\subsection{Fixed-phase as a special case of the fixed-node}
 We note that the fixed-phase approximation is a generalization of the more familiar fixed-node approximation for a real-valued $\Psi_T(\mathbf{R})$. Let us show that explicitly. 
We denote the nodes of $\Psi_T$ as 
\begin{equation}
\Gamma =\left\lbrace \mathbf{R}; \Psi_T(\mathbf{R})=0 \right\rbrace 
\end{equation}
Consider another wave function $\Psi_B(\mathbf{R})$ that is symmetric, normalizable and real. It should be also nonvanishing and positive in the domain of $\Psi_T(\mathbf{R})$. 
An appropriate prototype for $\Psi_B$ can be, for example, an approximation to the bosonic ground state of $H$. 
We construct a new complex trial function
\begin{equation}
\tilde\Psi_T = \Psi_T +i\varepsilon \Psi_B   
\end{equation}
The potential that is generated by the phase of 
$\tilde\Phi_T(\mathbf{R})$ is given by
\begin{equation}
V_{ph} = \frac{1}{2}|\nabla\tilde\Psi_T|^2  
=\frac{1}{2}\left|{\varepsilon \mathbf{h}\over \Psi_T^2+\varepsilon^2\Psi_B^2}\right|^2
\end{equation}
where
\begin{equation}
\mathbf{h}= \Psi_T \nabla \Psi_B - \Psi_B \nabla \Psi_T.    
\end{equation}
Away from the node $\Gamma$ the limit $\varepsilon \to 0$
produces $V_{ph}=0$ since then $\Psi_T^2 >0$. At the node $\Gamma$ the situation is a little bit more subtle. The key point is that the function $|\nabla\Psi_T|^2\ge 0$ is generically nonzero at the node (exceptions might possibly be non-analytical points of $\Psi_T$ due to interaction singularities, which are, however, of zero measure). Therefore taking the limit $\varepsilon \to 0$ we get
\begin{equation}
V_{ph}(\mathbf{R})= V_{\infty} \delta (\mathbf{R}-\mathbf{R}_{\Gamma})   
\end{equation}
where $\mathbf{R}_{\Gamma} \in \Gamma$ and  $V_{\infty}$ diverges as $\propto 1/\varepsilon^2$, therefore $V_{ph}$  enforces vanishing of any wave function at the node $\Gamma$. In this limit $V_{ph}$ become the fixed-node ``potential" that is more naturally understood as a boundary condition. 
The fixed-phase approximation is therefore more general than the fixed-node approximation. However, the accuracy of the method depends on the choice of the phase that nominally varies in the full configuration  space unlike in the fixed-node condition that applies only on the nodal subspace that is $(3N-1)$-dimensional for $N$ fermions in 3D space (ie, its codimension is 1).

\subsection{Importance sampling}
 If we try to solve equation (\ref{rho-eqn}) without modification, fluctuations in the weights due to the potentials will make the DMC implementation inefficient.
 We therefore apply an importance sampling transformation with a trial amplitude \cite{reynolds82}. If we denote $g(\mathbf{R},\tau) = \rho_T(\mathbf{R})\rho(\mathbf{R},\tau)$, equation (\ref{rho-eqn}) becomes 
\begin{equation}
    \begin{split}
	\label{importance-sampling} -\frac{\partial g(\mathbf{R},\tau)}{\partial \tau} = -\frac{1}{2}\nabla^2 g(\mathbf{R},\tau) + \nabla \cdot \left[ \mathbf{v}_D(\mathbf{R}) g(\mathbf{R},\tau)\right] \\+ \left[ E_L(\mathbf{R}) - E_T \right] g(\mathbf{R},\tau)
\end{split}
\end{equation}
where we have included also an energy offset $E_T$. 
The importance sampling introduces two new terms, namely a drift velocity 
\begin{equation}
\mathbf{v}_D(\mathbf{R}) = \nabla \ln  \rho_T(\mathbf{R}) = \rho_T^{-1}(\mathbf{R}) \nabla \rho_T(\mathbf{R})
\end{equation}
and the local energy 
\begin{equation}
  \label{eqn:local_energy}  E_L(\mathbf{R})  = \rho_T^{-1}(\mathbf{R}) \left[ -\frac{1}{2} \nabla^2 + V + \frac{1}{2}\left| \nabla \Phi_T(\mathbf{R})\right|^2 \right] \rho_T(R)
\end{equation}
For later purposes we can simplify the evolution equation by denoting the dynamical part of the operator 
acting on the function $g$ as $H^{\rm drift}_{\mathbf R}$ so that we can write
\begin{equation}
\label{drift}
-\frac{\partial g(\mathbf{R},\tau)}{\partial \tau} =
[H^{\rm drift}_{\mathbf R} +E_{L}(\mathbf{R})-E_T] g(\mathbf{R},\tau)
\end{equation}
One can find straightforward formulas for the drift, potential generated by the phase and local energy by using the gradient and laplacian of $\Psi_T$. Clearly, we have 
\begin{equation}
  \label{drift2}  \nabla \rho_Te^{i\Phi_T}  = e^{i\Phi_T}\nabla\rho_T + \Psi_T (i\nabla\Phi_T)
\end{equation}
which implies
\begin{equation}
\nabla\Phi_T =\rm{Im} (\Psi_T^{*}\nabla\Psi_T )/\rho_T^2
\end{equation}
\begin{equation}
\nabla\ln\rho_T =\rm{Re} (\Psi_T^{*}\nabla\Psi_T)/\rho_T^2
\end{equation}
Similarly for the laplacian we write
\begin{multline}
\nabla^2\Psi_T=\nabla^2[\rho_Te^{i\Phi_T}] = e^{i\Phi_T}\nabla^2\rho_T -\rho_T e^{i\Phi_T}(\nabla\Phi_T)^2\\
+2i e^{i\Phi_T}(\nabla\rho_T\cdot \nabla \Phi_T)+
i\rho_Te^{i\Phi_T}
(\nabla^2\Phi_T)
\end{multline}
so that the real contribution can be further arranged as
\begin{equation}
\rm{Re}[\Psi_T^{*}T_{kin}\Psi_T]= \rho_T(-1/2)\nabla^2\rho_T+
\rho_T^2(1/2)(\nabla\Phi_T)^2
\end{equation} 
where $T_{kin}=-(1/2)\sum_i\nabla^2_i$.
Therefore we can write 
\begin{equation}
\rm{Re}[\Psi_T^{*}T_{kin}\Psi_T]/\rho_T^2= \rho_T^{-1} T_{kin}\rho_T+
(1/2)(\nabla\Phi_T)^2
\end{equation}
and the local energy is then
\begin{equation}
E_L (\mathbf{R})= Re[\Psi_T^{*}T_{kin}\Psi_T]/\rho_T^2 + V 
\end{equation} 
Since gradient and laplacian of $\Psi_T$ are routinely calculated in DMC, by using the above formulas the needed quantities can be evaluated straightforwardly.

Rewriting equation (\ref{importance-sampling}) in integral form yields
\begin{equation}
    g(\mathbf{R}',t+\tau) = \int \dd \mathbf{R} \; \rho_T(\mathbf{R}')G(\mathbf{R}' \leftarrow \mathbf{R},\tau)\rho_T^{-1}(\mathbf{R})g(\mathbf{R},t)
\end{equation}
The Green's function for this process is of the exact same form as the in FNDMC, 
\begin{multline}
    \label{eqn:greens_fn} G(\mathbf{R}' \leftarrow \mathbf{R}; \tau) \simeq (2 \pi \tau)^{-3N/2}\exp\left[ \frac{-\left|\mathbf{R}'-\mathbf{R}-\tau \mathbf{v}_D(\mathbf{R})\right|^2}{2\tau} \right] \\
    \times \exp \left[ -\frac{ \tau}{2}(E_L(\mathbf{R'})+E_L(\mathbf{R})-2E_T)\right]
\end{multline}
At this point, it is clear that the implementation of FPDMC proceeds in the exact same manner as FNDMC. The main difference is that the mixed-distribution is made from the {\em trial amplitude} rather than the trial wave function. Additionally, the local energy has an additional term from the trial phase. Since both amplitudes in $g(\mathbf{R},\tau)$ are positive-definite everywhere, any proposed move in the imaginary time evolution is accessible since there is no nodal surface (any incidental zeros of $\rho_T$ are at most codimension 2, ie, of zero measure, similarly to points in 3D space).
The accuracy of this method clearly depends on the accuracy of the trial phase. If the trial phase happens to be the exact phase, then the projected solution will be $g(\mathbf{R},\infty) \propto \rho_T(\mathbf{R})\rho_0(\mathbf{R},\infty)$ producing the ground state energy while the convergence towards the exact value scales with the square of the difference between the exact 
and approximate trial function.

\section{Spin Orbit Interactions}\label{section:spinorbit}
In this section, we introduce the relativistic part of the Hamiltonian for use in FPDMC. In \S \ref{section:arep_and_so}, we give a quick discussion of relativistic quantum mechanics in a 4-component formalism. We then discuss the reduction to a 2-component formalism with an effective Hamiltonian. This effective Hamiltonian approach uses pseudopotentials to replace the relativistic core-electrons with a suitable effective field for the valence electrons. We discuss the similarities and differences between the standard effective core potentials used in DMC. In \S \ref{section:upperbound}, we show that an upper bound can be obtained within the fixed-phase approximation for complex nonlocal operators like our spin-orbit Hamiltonian. 

\subsection{AREP and SO Operators}\label{section:arep_and_so}
 For heavy atomic and molecular systems, bonding and spectral properties cannot be accurately predicted without the inclusion of scalar-relativistic and spin-orbit effects \cite{desclaux,lu}. 
In a relativistic treatment, one must begin with the approximation for the relativistic Hamiltonian \cite{grant,desclaux2}, known as the Dirac-Coulomb Hamiltonian, given as 
\begin{equation}
   \label{dirac_coulomb} H = \sum\limits_{i=1}^{N_e} \left[-ic(\bm{\alpha}\cdot\nabla)_i + \bm{\beta}c^2 \right] - \sum\limits_{i=1}^{N_e}\sum\limits_{I=1}^{N_{ion}}\frac{Z_I}{r_{iI}} + \sum\limits_{i>j} \frac{1}{r_{ij}}
\end{equation}
where $\bm{\alpha}$ and $\bm{\beta}$ are defined through the Pauli matrices $\bm{\sigma}$ and identity $I_2$,
\begin{equation}
\begin{array}{l r}
\bm{\alpha} = \left( \begin{array}{cc} 0 & \bm{\sigma} \\ \bm{\sigma} & 0 \end{array}\right),  & \bm{\beta} = \left( \begin{array}{cc} I_2 & 0 \\ 0 & I_2 \end{array} \right) 
\end{array}
\end{equation}
Here we have ignored the Breit interactions\cite{breit}, which gives rise to higher order retardation effects such as spin-other orbit, dipole interactions between two spins, and Fermi-contact interactions. This Hamiltonian, to order $1/c^2$, contains the dominant relativistic effects of the mass-velocity correction, the Darwin contribution to the $\ell = 0$ atomic level, and the spin-orbit interaction. The eigenfunctions of the Hamiltonian in equation (\ref{dirac_coulomb}) will be 4-component Dirac spinors. These spinors can be decomposed into large and small 2-component spinors, $\psi_L$ and $\psi_S$ respectively. Analysis of the individual $\psi_L$ and $\psi_S$ for all-electron systems indicate that $\psi_S$ is negligible in the valence region where chemical bonding is important \cite{desclaux3}. From this, to a reasonable approximation the valence electrons can be accurately described by 2-component spinors. Since the relativistic effects are strongest in the core region, valence electrons can be treated nonrelativistically subject to an effective field that mimics the repulsion of the core electrons \cite{lee}. This points to representing the relativistic effects on the valence electrons through an operator $W^{REP}$ that leads to the following
Hamiltonian
\begin{equation}
H = T_{kin}+ V + W^{REP}
\end{equation}
which contains only the valence electrons. The effective potential $W^{REP}$ for an electron $i$ from a given ion is typically expanded in the form
\begin{equation}
\label{rep}W^{REP}_i = \sum\limits_\ell \sum\limits_{j=|\ell-1/2|}^{\ell+1/2} \sum\limits_{m_j = -j}^j W^{REP}_{\ell j}(r_i) |\ell j m_j\rangle \langle \ell j m_j | 
\end{equation}
where $r_i$ is the electron-ion distance.
The effective operator $W^{REP}$ contains all of the relativistic effects from the core region and allows one to only consider the valence electrons, as is typically done in nonrelativistic calculations \cite{kahn}. This is very important for the application in DMC, where for all-electron systems the computational demands scale with the atomic number $Z$ as $\approx Z^6$, but scales more favorably as $N^{2-3}_{valence}$ when effective potentials remove the core electrons.  Therefore, relativistic QMC calculations can be done using the effective non-local pseudopotentials. 

The relativistic effective potential can be written in a different form that separates the relativistic effects into scalar relativistic and spin-orbit\cite{ermler}, namely $W^{REP} = W^{AREP} + W^{SO}$. The operator is semi-local; i.e. local in the relative distance to the nearest nucleus, but non-local in the solid angle for a given radius. The first term is spin-averaged core potential, which includes the effect of the mass-velocity, Darwin, averaged spin-orbit, and the effective field under which the valence electrons respond. The AREP term takes the form
\begin{multline}
W^{AREP}_i = W^{AREP}_{L} (r_i)  \\
+ \sum\limits_\ell^{L-1} \sum\limits_{m=-\ell}^{\ell} \left[W^{AREP}_\ell(r_i) - W^{AREP}_L(r_i) \right] |\ell m\rangle \langle \ell m | 
\end{multline}
where $W^{AREP}_L$ is the local part of the potential, and $W_\ell^{AREP}(r_i)$ is weighted average over the $j$ terms in equation (\ref{rep}), 
\begin{equation}
    W^{AREP}_\ell(r_i) = \frac{1}{2\ell+1} \left[ \ell W^{REP}_{\ell,\ell-1/2}(r_i) +(\ell+1)W^{REP}_{\ell,\ell+1/2}(r_i) \right]
\end{equation}
The spin-orbit interaction is included in the $W^{SO}$ operator and takes the form
\begin{multline}
    W^{SO}_i = s \cdot \sum\limits_{\ell=1}^L \frac{2}{2\ell+1}\Delta W_\ell^{SO}(r_i) 
   \\ 
\times    \sum\limits_{m=-\ell}^\ell \sum\limits_{m'=-\ell}^\ell |\ell m\rangle \langle \ell m | \ell | \ell m'\rangle\langle \ell m'|
\end{multline}
with the definition $\Delta W^{SO}_\ell(r_i) = W^{REP}_{\ell,\ell+1/2}(r_i) - W^{REP}_{\ell,\ell-1/2}(r_i)$.
The radial functions are expanded in gaussians in the same form as traditional nonrelativistic pseudopotentials, namely
\begin{equation}
    W^X_Y (r_i) = \frac{1}{r_i^2}\sum\limits_\alpha A_{\ell \alpha}r_i^{n_{\ell \alpha}} e^{-B_{\ell \alpha} r_i^2}
\end{equation}
where $X \in \{AREP,SO,REP\}$ and $Y\in\{\ell,L\}$ and $s$ is the spin. The parameters $A_{\ell,\alpha},$ $n_{\ell,\alpha}$ and $B_{\ell \alpha}$ for the $AREP$ and $SO$ terms have been developed by various groups including the Stuttgart-Cologne group \cite{stuttgart} and Clarkson University group \cite{clarkson}. 

In order to include relativistic effects into DMC, we must consider the action of the pseudopotential.  The action of the pseudopotential on the wave function will have real and imaginary parts, and thus the amplitude and phase equations are transformed to 
\begin{eqnarray}
\label{eqn:amp_with_pp}
    -\frac{\partial \rho}{\partial \tau} &=& \left[ -\frac{1}{2} \nabla^2 + V + \frac{1}{2}\left|\nabla \Phi\right|^2 + W^{Re}\right] \rho \\
     -\frac{\partial \Phi}{\partial \tau} &=& \left[ -\frac{1}{2} \nabla^2+\frac{\nabla \rho \cdot\nabla}{\rho} + W^{Im} \right] \Phi
\end{eqnarray}
with
\begin{equation}
    W^{Re/Im} = \textrm{Re/Im}\left[ \frac{W^{REP}\Psi}{\Psi}\right]
\end{equation}
The imaginary part describes the phase flux determined both by 
$\rho$ and $W^{Im}$. The real part is the eigenvalue equation that
provides the total energy eigenvalue of the system. 
Since we do not know the exact phase or the exact wave function to determine $W^{Re/Im}$, we invoke the fixed-phase approximation as discussed earlier as well as the localization approximation used in many conventional DMC calculations \cite{mitas91}. This projects the pseudopotential onto the trial wave function $\Psi_T$ as
\begin{equation}
    W^{Re} \rightarrow W^{Re}_T  = \textrm{Re}\left[ \frac{W^{REP}\Psi_T}{\Psi_T}\right]
\end{equation}

Note that the localization approximation eliminates the fundamental difficulty of the nonlocal operator that, 
in general, leads to introduction of another type of fermion sign problem, even for a single electron. 
This is easy to see by considering the matrix elements of the pseudopotential. If we denote the configuration space of space ${\bf r}_i$ and spin $s_i$ coordinates
\begin{equation}
\mathbf{X}=({\bf R},{\bf S}) =
({\bf r}_1, ...,{\bf r}_N,  s_1, ..., s_N)
\end{equation}
the matrix element that enters the Green's function after a Trotter expansion can be written as
\begin{equation}
    \begin{split}
	\label{eqn:nonlocal_expansion}
    \langle \mathbf{X}'|\exp(-\tau W) | \mathbf{X} \rangle
    =\delta(\mathbf{X}'-\mathbf{X})\\
   -\tau \langle \mathbf{X}'|W | \mathbf{X} \rangle +{\cal O}(\tau^2)
\end{split}
\end{equation}
The key problem lies with the matrix elements 
$\langle \mathbf{X}' |  W  | \mathbf{X} \rangle$
that, in general, can have a complicated sign structure and thus generate negative or complex values; obviously, this is also true without the spin-orbit terms.  
The locality approximation eliminates this problem. However, it generates a bias that vanishes quadratically with the error in the trial wave function and it also does not guarantee the variational property with regard
to the original Hamiltonian\cite{mitas91}.
For real valued wave functions, the variational property can be recovered \cite{tenHaaf} using the T-moves algorithm \cite{casula06}. In the next section we show that this upper bound can also be obtained for the complex pseudopotentials and wave functions. 

\subsection{Variational property of the fixed-phase method for nonlocal, complex, Hermitian operators }\label{section:upperbound}
In the following, we present a generalization of the proof of the upper-bound for nonlocal operators and real wave functions\cite{tenHaaf} with combined sampling and localization 
projection given above that enables to recover the upper bound property. Here we will show it for more general nonlocal Hermitian operators and complex wave functions. This proof enables us to be able to apply the 
T-moves technique \cite{casula06} in this more general setting. In the proof we will follow rather closely the original arguments \cite{tenHaaf} that will be generalized at a few important points. 

For a nonlocal operator $W$ sign changes arise when the following condition is fulfilled  
\begin{equation}
    \label{sign-condition}
    \textrm{Re}\left[ \Psi_T^*(\mathbf{X})\Psi_T(\mathbf{X}'') \langle \mathbf{X} | W | \mathbf{X}'' \rangle\right] > 0
\end{equation}
as is clear from considerations of Eq. \ref{eqn:nonlocal_expansion} in the importance sampling Green's function.
Note that we can express the matrix elements of arbitrary Hermitian operator as $\langle \mathbf{X} | W | \mathbf{X}''\rangle = w(\mathbf{X},\mathbf{X}'')e^{i \gamma(\mathbf{X},\mathbf{X}'')}$ where $w(\mathbf{X},\mathbf{X}'')$ is symmetric in 
$ \mathbf{X} \leftrightarrow \mathbf{X}''$ and positive-definite. Then $\gamma(\mathbf{X},\mathbf{X}'') = -\gamma(\mathbf{X}'',\mathbf{X})$ as must be the case when  $W$ is Hermitian, i.e., $W=W^\dagger$. We write the trial wave function as $\Psi_T(\mathbf{X}) = \rho_T(\mathbf{X}) e^{i \Phi_T(\mathbf{X})}$. Denoting 
\begin{equation}
\alpha(\mathbf{X},\mathbf{X}'') = \Phi_T(\mathbf{X}'')-\Phi_T(\mathbf{X}),
\end{equation}
 condition (\ref{sign-condition}) becomes
\begin{equation}
    \textrm{Re}\left[ w(\mathbf{X},\mathbf{X}'')e^{i\gamma(\mathbf{X},\mathbf{X}'')}\rho_T(\mathbf{X})\rho_T(\mathbf{X}'')e^{i \alpha(\mathbf{X},\mathbf{X}'')}\right] > 0
\end{equation}
which reduces to 
\begin{equation}
    \cos(\alpha(\mathbf{X},\mathbf{X}'')+\gamma(\mathbf{X},\mathbf{X}'')) > 0
\end{equation}
since $w(\mathbf{X},\mathbf{X}'')$, $\rho_T(\mathbf{X})$, and $\rho_T(\mathbf{X}'')$ are positive-definite. Although $\alpha$ and $\gamma$ are anti-symmetric, $\cos(\alpha+\gamma)$ is a symmetric function of $\mathbf{X}$ and $\mathbf{X}''$. Following \cite{tenHaaf,casula06}, we construct an effective Hamiltonian 
\begin{multline}
\langle \mathbf{X} | H_{eff} | \mathbf{X}' \rangle \\ 
\quad \\
=\left\{ \begin{array}{l l} \langle \mathbf{X}|H|\mathbf{X}' \rangle & , \mathbf{X} \ne \mathbf{X}' \textrm{ and } \cos(\alpha+\gamma)<0 \\ 
0 & ,\mathbf{X}\ne \mathbf{X}' \textrm{ and } \cos(\alpha+\gamma) > 0 \\
\langle \mathbf{X} | H + V_{sf}| \mathbf{X}' \rangle &, \mathbf{X}=\mathbf{X}' 
\end{array}
\right.
\end{multline}
where $V_{sf}$ is the {\em sign-flip} potential defined as 
\begin{equation}
   \langle \mathbf{X} | V_{sf} | \mathbf{X} \rangle = \int\limits_{\cos(\alpha+\gamma) > 0 } \dd \mathbf{X}' \; \langle \mathbf{X}|W|\mathbf{X}'\rangle \frac{\Psi_T(\mathbf{X}')}{\Psi_T(\mathbf{X})}
\end{equation}
We want to show that $H_{eff}$ produces an upper bound for the original Hamiltonian. We begin with any state with the {\em same phase} as the trial wave function, namely
\begin{equation}
    |\Psi \rangle = \int \dd \mathbf{X} \; \Psi(\mathbf{X}) |\mathbf{X} \rangle = \int \; \dd \mathbf{X} \;\rho(\mathbf{X}) e^{i \Phi_T(\mathbf{X})} |\mathbf{X}\rangle
\end{equation}
The discrepancy of $H_{eff}$ and $H$ with this state is
\begin{equation}
\begin{split}
    \Delta E &= \langle \Psi | H_{eff} - H | \Psi \rangle \\
             &= \langle \Psi | V_{sf} - H_{sf} | \Psi \rangle
             \end{split}
\end{equation}
Rewriting this in configuration space, we obtain
\begin{multline}
    \Delta E = \int \dd \mathbf{X} \;\Psi^*(\mathbf{X}) \Big[ \langle \mathbf{X}|V_{sf}| \mathbf{X} \rangle \Psi(\mathbf{X})
    \\
    -\int \dd \mathbf{X}' \langle \mathbf{X} | H_{sf} | \mathbf{X}' \rangle \Psi(\mathbf{X}') \Big]
\end{multline}
Rewriting this over the terms that generate sign-flips, we have
\begin{multline}
    \Delta E = \int\dd \mathbf{X} \int\limits_{sf} \dd \mathbf{X}' \left| \Psi(\mathbf{X}) \right|^2 \langle \mathbf{X} | W | \mathbf{X}' \rangle \frac{\Psi_T(\mathbf{X}')}{\Psi_T(\mathbf{X})} 
   \\ -\langle \mathbf{X}|W|\mathbf{X}'\rangle \Psi^*(\mathbf{X}) \Psi(\mathbf{X}')
\end{multline}
Denoting $h = \rho_T(\mathbf{X}')/\rho_T(\mathbf{X})$, we see that 
\begin{multline}
    \Delta E = \iint\limits_{\Omega} \dd \mathbf{X} \dd \mathbf{X}' \; w(\mathbf{X},\mathbf{X}')\cos(\alpha(\mathbf{X},\mathbf{X}')+\gamma(\mathbf{X},\mathbf{X}')) \\
\times    \left[ h\rho^2(\mathbf{X}) + h^{-1} \rho^2(\mathbf{X}') - 2 \rho(\mathbf{X})\rho(\mathbf{X}')\right]
\end{multline}
where $\Omega=\{{\bf X,X'}; \cos(\alpha+\gamma) > 0 \}$. 
If we simplify this expression once more, we see that $\Delta E$ becomes
\begin{multline}
 \Delta E =   \iint\limits_{\cos(\alpha+\gamma)>0} \dd \mathbf{X} \dd \mathbf{X}' w(\mathbf{X},\mathbf{X}') \\
    \times \cos(\alpha(\mathbf{X},\mathbf{X}')+\gamma(\mathbf{X},\mathbf{X}')) \frac{\left[h\rho(\mathbf{X})-\rho(\mathbf{X}')\right]^2}{h}
  \geq 0  
\end{multline}
which is clearly positive everywhere, since the integration is over the region where $\cos(\alpha(\mathbf{X},\mathbf{X}')+\gamma(\mathbf{X},\mathbf{X}'))$ is positive. Thus, the effective Hamiltonian produces an upper bound for $H$ and recovers the variational property. Note that this point - that the approximation makes sense only in the fixed-node framework - has been emphasized in the original paper on the localization approximation \cite{mitas91}.
The upper bound can be recovered in this framework by implementation of the so-called $T-$moves algorithm \cite{casula06}. Note that even in this algorithm the fixed-node/phase condition is
of key importance, the T-moves can recover only the best possible energy within the given constraint. 


\section{Spin Representation and Sampling}\label{section:spin}
We discuss how spins can be treated as a quantum variable. In \S \ref{section:spinrep}, we discuss our choice for a continuous and overcomplete representation of the spin variable. Once we have a representation for the spin variable, we discuss the form of the one-particle spinors and trial wave functions which couple the spin and spatial degrees of freedom in \S \ref{section:wf}. We discuss the evaluation of the pseudopotential with this spin representation in \S \ref{section:evalPP}. Lastly, we discuss our choice of sampling spin degrees of freedom and how this modifies the Green's function in FPDMC in \S \ref{section:spinsampling}. 

\subsection{Spin Representations}\label{section:spinrep}

Let us denote one-particle spinors as
\begin{equation}
\label{eqn:spinor}
\chi ({\bf r},s)=\alpha\varphi^{\uparrow}({\bf r})\chi^{\uparrow}(s)
 +\beta\varphi^{\downarrow}({\bf r})\chi^{\downarrow}(s)
\end{equation}
where $s$ is the spin projection coordinate on 
the $z-$axis.  
 In its usual minimal representation
the spin variable have
discrete values  $s=\pm 1/2$ so that for $S_z$ eigenstates 
 $\chi^{\uparrow}(1/2)=\chi^{\downarrow}(-1/2)=1$,
$\chi^{\downarrow}(1/2)=\chi^{\uparrow}(-1/2)=0$. 
 The evaluation of any expectation $\langle {\cal B}\rangle$ for a variational wave function  $\Psi_{var}({\bf R,S})=\Psi_{var}({\bf X})$ includes  spatial integrations as well as
 summation over $2^N$ spin configurations space of $(-1/2,1/2)^N$ 
 \begin{equation}
 \label{eqn:expectation_value}
 \langle {\cal B} \rangle_{var}  = \frac{ \int d{{\bf R}}\sum_{\bf S} \Psi_{var}^{*}{\cal B}\Psi_{var}}
{  \int d{{\bf R}}\sum_{\bf S} \Psi_{var}^{*}\Psi_{var}}=
\frac{ \int d{{\bf X}} \Psi_{var}^{*}{\cal B}\Psi_{var}}
{  \int d{{\bf X}} \Psi_{var}^{*}\Psi_{var}}
 \end{equation}
 assuming we have $N$ fermions. This can be recast
 as sampling according to the positive density $w({\bf R,S})=|\Psi_{var}({\bf R,S})|^2$
 \begin{equation}
 \langle {\cal B} \rangle_{var}  = \frac{ \int d{\bf X} w({\bf X})(\Psi_{var}^{*}({\bf X}))^{-1}{\cal B}\Psi_{var}({\bf X})}
{  \int d{\bf X} w({\bf X}) }
 \end{equation}
 
 This expression can be implemented in the variational Monte Carlo (VMC) as one can simply add the sampling of the spin configurations 
 to the discrete sampling of the spatial coordinates. 

However, generalization to projection methods is more complicated. 
Note that any change of the discrete spin coordinate(s) will lead to ``jumps'' in the stochastic path. These jumps can cause the local energy fluctuations to increase substantially and that could possibly compromise the utility and efficiency of the method. Since we are employing a diffusion-drift sampling process in imaginary time
with weights that include local energy in the exponential, any large fluctuations would make a reliable estimate of the expectations difficult to obtain, especially if we increase the system size. 
Another strategy would be to sum
over all of the spin configurations for every spatial step.  However, this has an exponential scaling so that for large systems this is intractable. 

One possibility how to address this obstacle is to make the spin configuration space compact and continuous, which allows for smooth sampling \cite{kevin}. We can choose an {\em overcomplete} spin representation  through the utilization of a 1D ring (or $S^1$) lowest pair of degenerate eigenstates as follows:
\begin{equation}
\begin{array}{l r} \langle s_j | \chi^\uparrow \rangle = e^{i s_j},\;\;\; & \langle s_j | \chi^\downarrow \rangle = e^{-i s_j}  \ \end{array}
\end{equation}
where the spin variable $s_j \in [0,2\pi)$. This implies the normalization condition for two arbitrary spin states
\begin{equation}
\langle \chi^\alpha | \chi^\beta \rangle = \int_0^{2\pi} \frac{\dd s}{2\pi} \; \langle \chi^\alpha | s\rangle  \langle s | \chi^\beta \rangle = \delta_{\alpha\beta}
\end{equation}

As the simplest illustration, consider an arbitrary one-electron spinor of the form $| \chi \rangle = a |\chi^\uparrow\rangle + b | \chi^\downarrow\rangle$. The expectation value of the $S_x = \frac{1}{2} \left[ |\chi^\uparrow\rangle \langle \chi^\downarrow| + |\chi^\downarrow\rangle\langle \chi^\uparrow|\right]$ operator with the spinor is clearly $\langle \chi | S_x | \chi \rangle = ab$. If we now consider this expectation value in a VMC formulation, we have the following expectation value
\begin{equation}
    \langle \chi | S_x | \chi \rangle = \int \dd s \; |\chi(s)|^2 E_L(s)
\end{equation}
where $\chi(s) = a e^{is}+b e^{-is}$ and the local energy is $E_L(s) = \chi^{-1}(s)S_x \chi(s) = 1/2(ae^{-is}+be^{is})/(ae^{is}+be^{-is})$. Plugging in, this yields the expectation value $ab$ as expected, where we sample the distribution $|\chi(s)|^2$ and evaluate the average of the local energy.

The introduced representation has several important consequences. First, 
it enables to define a continuous path for the evolving sampling 
points (walkers) and therefore all the associated quantities 
along the path are smooth
by definition. Second, the spin coordinate space that is introduced has some desireable properties, namely is it is compact and the  interval $(0,2\pi)$ can be sampled rapidly. The harmonic functions  have minimal curvature and are complex so that no additional (artificial) node  created, ie, so that the formulation fits  the fixed-phase formulation. Third, note that unlike 
discrete coordinates that switch-on and -off the up and down components of the spinor, the spin functions are always somewhere ``in between" due to the fact that they are weighted by complex values with unit modulus.
In effect, they introduce a complex weighted spinors that for many-spins hedge the average effect of the spin summations. This will prove important at sampling the spin coordinates as explained below.

\subsection{Trial Wave Functions}\label{section:wf}

In FNDMC calculations without spin terms in the Hamiltonian,
the electrons can be labeled as $N_\uparrow$ spin-up and $N_\downarrow$ spin-down ones and these labels remain static. 
This is due to the fact that spins commute with the Hamiltonian, implying that both the total and individuals spins are conserved.  It can be shown \cite{qmcrev} that expectation values can be then calculated using spatial only averaging with configuration space of $\mathbf{R} \in \mathbb{R}^{3N}$. Consequently, the trial wave functions are typically constructed as bipartitioned spin-up and 
-down Slater determinant(s) built from one-particle orbitals obtained from Hartree-Fock, post-Hartree-Fock or DFT methods  
\begin{equation}
    \Psi_T(\mathbf{R}) = e^{U(\mathbf{R})} \sum_m c_m \textrm{det}_m^{\uparrow}\left[\phi_i(\mathbf{r}_k)\right]
    \textrm{det}_m^{\downarrow}\left[\phi_j(\mathbf{r}_l)\right]
\end{equation}
The particle correlations are explicitly approximated by the Jastrow factor given as  
\begin{equation}
    U(\mathbf{R}) = \sum\limits_{iI}U_1(r_{iI}) _+ \sum\limits_{i\ne j} U_2(r_{ij}) + \sum_{I,i\ne j} U_3(r_{iI},r_{jI},r_{ij})
\end{equation}
 where we have one-, two-, and three-body terms ($U_1(r_{i,I}), \, U_2(r_{ij}),$ and $U_3(r_{iI},r_{jI},r_{ij})$ respectively), that 
describe electron-ion, electron-electron, etc, correlations.

With our choice of spin representation in \S \ref{section:spinrep}, we have a wave function that lives in a configuration space $\mathbf{X} = \{ (\mathbf{r}_1,s_1), \ldots, (\mathbf{r}_N,s_N)\} \in \mathbb{R}^{3N} \times [0,2\pi)^{N}$.
We write the trial wave function as 
\begin{equation}
    \Psi_T(\mathbf{X}) = e^{U(\mathbf{R})} \sum\limits_\alpha c_\alpha \textrm{det}_\alpha \left[ \ldots, \chi_i(\mathbf{r}_k,s_k) ,\ldots \right]
\end{equation}
where $\{ \chi_i(\mathbf{r},s) \}$ are one-particle spinors. In general, each spinor has different spatial dependence for the up and down spin components, namely
\begin{equation}
    \chi(\mathbf{r},s) = a \varphi^{\uparrow}(\mathbf{r})e^{i s} + b \varphi^{\downarrow}(\mathbf{r})e^{-is}
\end{equation}
Each spatial function $\varphi^{\uparrow(\downarrow)}$ is expanded in appropriate basis functions (for example, gaussian type orbitals or plane waves). With regards to the Jastrow factor, we use the same form as described above. Seemingly, every electron should be treated as having the ``same'' spin since there is only {\em one} determinant of spinors, rather than the spin-like and spin-unlike distinction in Jastrow forms employed in conventional calculations. This would imply that the cusp should correspond to the like-spin value since the determinant vanishes at the two-electron coincidence point. However, spatial coincidence configurations are of zero measure 
with regard to coincidence at the full space-spin configurations space. Clearly, the differences in spin coordinates
make the determinant, in general, nonvanishing even when the spatial coordinates of two electrons coincide.
Therefore the more appropriate is 
the unlike spins cusp condition \cite{ceperley,ceperley_alder} 
\begin{equation}
    \left. \frac{\dd U_2(r_{ij})}{\dd r_{ij}} \right|_{r_{ij}=0} = 1/2
\end{equation} 
We also note that
the precise cusp value has only a marginal impact on the results 
since it really affects only a very small part of the configuration space.
Much more substantial effect comes from the shape of the Jastrow correlations for $r_{ij} > 0.1 - 0.2 $ Bohr, namely, at medium- and long-range distances. These are the ranges of distances where  correlations affect one- and two-electron pair densities very significantly over a sizable part of the configuration space. The 
accuracy in these regions has then important consequences for both minimization of energy fluctutations as well as for accurate projections and minimization of the localization bias.

In the limit of vanishing spin-orbit interaction
the single-reference spinor determinant (regardless of the chosen representation) should  
simplify to the product of spin-up and -down determinants. This is true also for our trial wave function. Let us consider $N$ occupied spinors that can be grouped as
as $N/2$ Kramer's pairs (for simplicity assuming $N$ to be even). 
 We can write the Kramer's pair as
\begin{eqnarray}
\chi^+=(\varphi+\Delta\varphi)\chi^{\uparrow}+
(\varphi-\Delta\varphi)\chi^{\downarrow} \\
\chi^-=(\varphi-\Delta\varphi)\chi^{\uparrow}-
(\varphi+\Delta\varphi)\chi^{\downarrow}
\end{eqnarray}
where the $\Delta\varphi$ is the spin-orbit induced splitting 
of the spatial orbital $\varphi$. We sketch a block of the first four rows 
from the corresponding Slater determinant as given by
\begin{equation}
{\rm det} \left[\begin{matrix}
\chi_1^+(1), \chi_1^+(2), \chi_1^+(3), \chi_1^+(4),... \\
\chi_1^-(1),  \chi_1^-(2), \chi_1^-(3), \chi_1^-(4),... \\
\chi_2^+(1), \chi_2^+(2), \chi_2^+(3), \chi_2^+(4),... \\
\chi_2^-(1),  \chi_2^-(2), \chi_2^-(3), \chi_2^-(4),... \\

...
\end{matrix}\right].
\end{equation}
Let all the spin variables $\{s_i\}$ have distinct values and the spin orbit
splitting $\Delta\varphi \to 0$. After some linear rearragements we can write the matrix as sketched for the first four rows 
\begin{equation}
{\rm det} \left[\begin{matrix}
\varphi_1(1), 0,\varphi_1(3),0, ... \\
0, \varphi_1(2), 0, \varphi_1(4), ...\\
\varphi_2(1), 0,\varphi_2(3),0, ... \\
0, \varphi_2(2), 0, \varphi_2(4), ...\\
... \\
\end{matrix}\right]
\end{equation}
up to a common complex prefactor. After reshuffling rows and columns, the single determinant of spinors factorizes into the product of two determinants with spin-up and -down particles. Generalization to odd $N$ with un unpaired spinor is straightforward.

For the sake of completeness we note that the simplest trial function based on a {\em pair spinor} orbital
$\chi_{pair}({\bf r}_i,s_i,{\bf r}_j,s_j)$ written as an
antisymmetrized product 
of distinct pairs of particles 
results in a pfaffian 
\begin{equation}
\Psi_T({\bf R,S})
= {\rm pf} [\chi_{pair}({\bf r}_i,s_i,{\bf r}_j,s_j)]\exp[U({\bf R})].
\end{equation}
Obviously, the pair orbital itself is antisymmetric since 
the pfaffian is defined for a skew-symmetric matrix so that
$\chi_{pair}({\bf r}_i,s_i,{\bf r}_j,s_j)=-\chi_{pair}({\bf r}_j,s_j,{\bf r}_i,s_i)$, as 
explained previously \cite{bajdich}. Note also that for odd number 
of electrons the skew symmetric matrix can be expanded by an unpaired row and column with an unpaired spinor so that the resulting matrix is of even dimension. Therefore systems with odd number of electrons can be described by the corresponding pfaffian \cite{acta} as well  
(without boosting the matrix by another row and column the
pfaffian of matrix with odd dimensions would vanishes identically).

\subsection{Evaluation of the Pseudopotential and Importance Sampling }\label{section:evalPP}
The evaluation of the term Re$\left[\Psi_T^{-1}W_T^{REP} \Psi_T\right]$ is similar to the evaluation done in standard QMC calculations. Consider electron $i$ and nucleus $I$ whose relative distance is $r_{iI}$. We will need to calculate the contribution
\begin{multline}
\label{equation:ecp}
    \frac{W_{T,(iI)}^{REP}\Psi_T}{\Psi_T} = \sum\limits_{\ell,j}W^{REP}_{\ell,j}(r_{iI}) \int \dd \Omega'_{iI} \int\dd s'_i \\
    \times \sum\limits_m \langle \Omega_{iI} s_i |\ell j m\rangle \langle \ell j m | \Omega'_{iI} s'_i\rangle \\
    \times \frac{\Psi_T((\mathbf{r}_1,s_1), \ldots,(\mathbf{r}'_i,s'_i),\ldots,(\mathbf{r}_N,s_N)) }{ \Psi_T((\mathbf{r}_1,s_1, \ldots, (\mathbf{r}_i,s_i),\ldots, (\mathbf{r}_N,s_N)) }
\end{multline}
where ${\bf r}_{iI}={\bf r}_i - {\bf r}_I$, 
$\Omega,\Omega'$ are corresponding solid angles, and $\mathbf{r}_i' = (r_i,\Omega_i')$, while the integral in $s'$ is over the spin degree of freedom in a given representation. 
One advantage of the projection used in localization approximation is that the integration over the spins can be done explicitly and exactly for each determinant in the trial function expansion. Of course, this is true only for the case
of the Jastrow factor factor being spin independent as is our choice here.

To illustrate this, let us consider a trial wave function which is built from a single determinant. Focusing on an individual electron $i$ near a nucleus $I$ as we have in equation (\ref{equation:ecp}), we will need the ratio of the wave function evaluated at $(\mathbf{r}'_i,s'_i)$ to the original wave function. This can be written as
\begin{equation}
    \frac{ \textrm{det}\left[\ldots,\chi_\alpha(\mathbf{r}_i',s'_i),\ldots\right]}{\textrm{det} \left[ \ldots, \chi_\alpha(\mathbf{r}_i,s_i),\ldots\right]} = \sum\limits_\alpha C_{\alpha,i}(\mathbf{r}_i,s_i)\chi_\alpha(\mathbf{r}'_i,s'_i)
\end{equation}
where $C_{{\alpha} ,i}(\mathbf{r}_i,s_i)$ are the matrix elements of the inverse transpose of the Slater matrix, where $\alpha$ labels the spinors and $i$ labels the electron. Plugging this into equation (\ref{equation:ecp}), we obtain
\begin{eqnarray}
\frac{W_{T,(iI)}^{REP}\Psi_T}{\Psi_T} = \sum\limits_{\ell,j}W^{REP}_{\ell,j}(r_{iI}) \sum\limits_\alpha C_{\alpha,i}(\mathbf{r}_i,s_i) \nonumber \\
\int \dd \Omega'_{iI} \int\dd s'_i \sum\limits_m \langle \Omega_{iI} s_i |\ell j m\rangle \langle \ell j m | \Omega'_{iI} s'_i\rangle \chi_\alpha(\mathbf{r}'_i,s'_i)
\end{eqnarray}
Focusing on an individual $\ell, j$ element in the summation, we see
\begin{multline}
W^{REP}_{\ell,j}(r_{iI})\sum_\alpha C_{\alpha,i}(\mathbf{r_i},s_i) \int \dd \Omega'_{iI} \int \dd s_i' \sum_m  \\
\times\langle \Omega_{iI} s_i |\ell j m\rangle \langle \ell j m | \Omega'_{iI} s'_i\rangle \left[a \phi_\alpha^\uparrow(\mathbf{r}')\chi^\uparrow(s') + b \phi_\alpha^\downarrow(\mathbf{r}')\chi^\downarrow(s_i') \right]
\end{multline}
where we have expanded the individual spinor $\chi_\alpha$ into its spin and spatial functions. 
We can simplify the previous expression by defining two functions $A_{\ell,j}$ and $B_{\ell,J}$ as 
\begin{eqnarray}
A_{\ell,j}(\Omega_{iI},\Omega_{iI}') &=& \int \dd s' \sum_m \langle \Omega_{iI} s_i |\ell j m\rangle \langle \ell j m | \Omega'_{iI} s'_i\rangle \chi^\uparrow(s'_i) \nonumber \\
B_{\ell,j}(\Omega_{iI},\Omega_{iI}') &=& \int \dd s' \sum_m \langle \Omega_{iI} s_i |\ell j m\rangle \langle \ell j m | \Omega'_{iI} s'_i\rangle \chi^\downarrow(s'_i) \nonumber \\
\end{eqnarray}
such that the individual $\ell,j$ element can be written as
\begin{multline}
W_{\ell,j}^{REP}(r_{iI})\sum\limits_{\alpha,i}C_{\alpha,i}(\mathbf{r}_i,s_i) \\
\times \int \dd \Omega_{iI}'\left[ a \phi^\uparrow(\mathbf{r}_i)A_{\ell,j}(\Omega_{iI},\Omega_{iI}') + b \phi^\downarrow(\mathbf{r}'_i)B_{\ell,j}(\Omega_{iI},\Omega_{iI}') \right] 
\end{multline}
Note that the spin integration has been eliminated, and we are only left with a integral over the solid angle $\Omega_{iI}'$, which is carried out numerically using standard techniques
as is the case of spatial-only nonlocality \cite{mitas91}. The terms $\langle \Omega s | \ell j m \rangle$ are spin-spherical harmonics with $j = \ell \pm 1/2$, and can be written in our spin representation as   
\begin{multline}
    \mathcal{Y}^{\ell}_{\ell+1/2,m}(\Omega,s) = \langle \Omega \, s | \ell,\,\ell+1/2, \, m\rangle  \\
    =\sqrt{\frac{\ell+m+1/2}{2\ell+1}} Y_{\ell, m-1/2}(\Omega)e^{is} \\
    + \sqrt{\frac{\ell - m + 1/2}{2\ell+1}}Y_{\ell,m+1/2}(\Omega)e^{-is}
\end{multline}
\begin{multline}
    \mathcal{Y}^{\ell}_{\ell-1/2,m}(\Omega,s)=\langle \Omega\, s | \ell, \, \ell-1/2, \, m\rangle \\
    = -\sqrt{\frac{\ell-m+1/2}{2\ell+1}}Y_{\ell,m-1/2}(\Omega)e^{is}\\
    +\sqrt{\frac{\ell+m+1/2}{2\ell+1}}Y_{\ell,m+1/2}(\Omega)e^{-is}
\end{multline}
Generalization to the inclusion of a spin-free Jastrow and/or multiple Slater determinant wave functions is straightforward. 

In order to obtain the entire contribution of the pseudopotential, we simply sum over all of the electrons and ions and add the local contribution $W^{REP}_L$. This yields a total pseudopotential contribution of 
\begin{equation}
    \label{eqn:nonlocal_energy}
    W^{Re}_T = \textrm{Re}\left\{ \sum\limits_{i=1}^{N_e}\sum\limits_{I=1}^{N_I}\left[W_{L}(r_{iI}) + \frac{W^{REP}_{T,(iI)} \Psi_T}{\Psi_T}\right]\right\}
\end{equation}
which is added to the local energy in equation (\ref{eqn:local_energy}).


Once we are able to evaluate the nonlocal potential contribution in the localization approximation as $W^{Re}_T
=Re[\Psi_T^{-1}W^{REP}\Psi_T]$, so that it becomes a multiplicative {\em many-body} and $\Psi_T$-dependent potential, we can apply the importance sampling 
transforomation to the Eq. \ref{eqn:amp_with_pp}. Note that due to the continuous spin values the corresponding local energy is
a continuous and smooth function almost everywhere (the exceptions might be, possibly, zero measure configurations for which 
both the amplitude $\rho$ and the phase $\Phi$ vanish simultaneously). The evolution equation is therefore now solved 
for the product $g=\rho_T\rho_{FP}$ where $FP$ denotes 
the fixed-phase solution.

\subsection{Spin Sampling}\label{section:spinsampling}
In order to sample the spin variables smoothly within this spin representation, we add a spin ``kinetic'' energy and corresponding offset into the Hamiltonian for all $s_i$, $i \in \{ 1,2,...,N \}$ of the form
\begin{equation}
    T_i^s = -\frac{1}{2\mu_s}\left[ \frac{\partial^2}{\partial s_i^2} +1\right]
\end{equation}
such that $H \rightarrow H' = H+\sum_{i=1}^N T_i^s$. 
Consider the action of $T^s_i$ on an arbitrary one-particle spinor $\psi(\mathbf{r}_i,s_i) = \alpha \varphi^\uparrow(\mathbf{r}_i)e^{is_i}+\beta \varphi^\downarrow(\mathbf{r}_i)e^{-i s_i}$, where $\varphi^{\uparrow(\downarrow)}(\mathbf{r})$ are different different spatial orbitals for the spin-up and -down components. Clearly, $T_i^s \psi(\mathbf{r}_i,s_i) = 0$ due to the introduced
offset so that there is no contribution from the spin laplacian.  

At this point one has to consider whether to apply the importance sampling
transformation before or after adding this spin variables Hamiltonian. Let us carry out the transformation with the spin 
Hamiltonian included, assuming the fixed-phase approximation. Using the notation from Eq. 
(\ref{drift}) it can be written as 
\begin{eqnarray}
  -\frac{\partial g(\mathbf{X})}{\partial \tau} = \left[ H_{\mathbf R}^{\rm drift}+E_{L}(\mathbf{X})+W_T^{Re} -E_T\right] g(\mathbf{X})\nonumber \\ 
   -\frac{1}{2\mu_s} \nabla^2_{\mathbf{S}} g(\mathbf{X})+  \frac{1}{\mu_s}\nabla_\mathbf{S}\cdot \left[ \mathbf{v}_D^\mathbf{S}(\mathbf{X}) g(\mathbf{X})\right]\nonumber \\ +\frac{1}{2\mu_s} \left| \nabla_\mathbf{S} \phi_T(\mathbf{X})\right|^2 g(\mathbf{X}) \nonumber\\ 
\end{eqnarray}
where we have written $\nabla_\mathbf{S} = (\frac{\partial}{\partial s_1},\ldots,\frac{\partial}{\partial s_N})$ and for the sake
of completeness we write down also the expression for the local energy 
\begin{equation}
E_{L}(\mathbf{X})= 
\textrm{Re}[\Psi_T^{*}(\mathbf{X})T_{kin}\Psi_T(\mathbf{X})]/\rho_T^2 + V 
\end{equation}
where $T_{kin}$ denotes the spatial kinetic energy.
Note that there
are formally three new terms generated by spin degrees of freedom: diffusion, drift term with velocity
\begin{equation}
    \mathbf{v}_D^\mathbf{S} (\mathbf{X}) = \nabla_\mathbf{S} \ln \rho_T(\mathbf{X})  = \rho_T^{-1}(\mathbf{X})\nabla_\mathbf{S} \rho_T(\mathbf{X})
\end{equation}
as well as contribution to the local energy from the spin gradient of the trial phase. 
We have introduced $H'$ only to be able to sample the spin variables $\mathbf{S}$ since in reality spin does not have any kinetic energy. Therefore we drop the contribution to the energy from the spin gradient of the trial phase as it creates an artificial contribution. This will be further discussed in \S \ref{section:errors}.
 
For completeness, the inclusion of the spin kinetic energy and offset modifies importance-sampling Green's function 
in equation (\ref{eqn:greens_fn}) to 
\begin{equation}
    \label{eqn:spin_space_greens} \widetilde{G}( \mathbf{X}'\leftarrow\mathbf{X};\tau) 
\simeq T_{\mathbf{X'},\mathbf{X}}
e^{-\tau[ E_L(\mathbf{X}) + E_L(\mathbf{X}')-2E_T]/2} 
\end{equation}
with
\begin{eqnarray}
    T_{\mathbf{X}',\mathbf{X}} &\propto& \exp\left[ \frac{-\left| \mathbf{R'}-\mathbf{R} - \tau \mathbf{v}_D^\mathbf{R}(\mathbf{R}) \right|^2}{2 \tau}\right] \nonumber \\ 
    &&\times \exp \left[ \frac{-\left| \mathbf{S}'-\mathbf{S}-\tau_s\mathbf{v}_D^{\mathbf{S}}(\mathbf{S})\right|^2}{2\tau_s}\right]
\end{eqnarray}
where we have introduced a spin time-step $\tau_s = \tau/\mu_s$.



\section{Time-step Errors and Approximations}\label{section:errors}
A potential source of error comes from the choice of spin time-step $\tau_s$ (or equivalently, the spin mass $\mu_s$) due to the complex representation on $S^1$. In its minimal representation, the expectation value for the energy is given by 
(\ref{eqn:expectation_value}), i.e. the entire $2^N$ spin configurations must be summed over at every step in the imaginary time evolution or one must sample the discrete spins causing ``jumps.'' In our continuous representation, the speed of the spin sampling can be chosen at our disposal due to the effective spin mass $\mu_s$. If we consider the $i$th walker, the expected root-mean-square displacement (rms) in coordinate space is
\begin{equation}
    r_i^{rms}(t) \propto \sqrt{t}
\end{equation}
neglecting the drift velocity. In spin space, however, the rms goes as
\begin{equation}
    s_i^{rms}(t) \propto \sqrt{\frac{t}{\mu_s}}
\end{equation}
By taking the limit $\mu_s \rightarrow 0$, the spin rms $s_i^{rms} \rightarrow \infty$, i.e. the spin has sampled its entire space. At $\mu_s = 0$, this is equivalent to summing over all spin configurations. 

\begin{figure}[!t]
\label{figure:spin_phase_grad}
\centering
\caption{Total Energy of the Pb ground state. The ``No Drift'' and ``Drift'' calculations are indistinguishable at this scale.}
\includegraphics[width=0.49\textwidth]{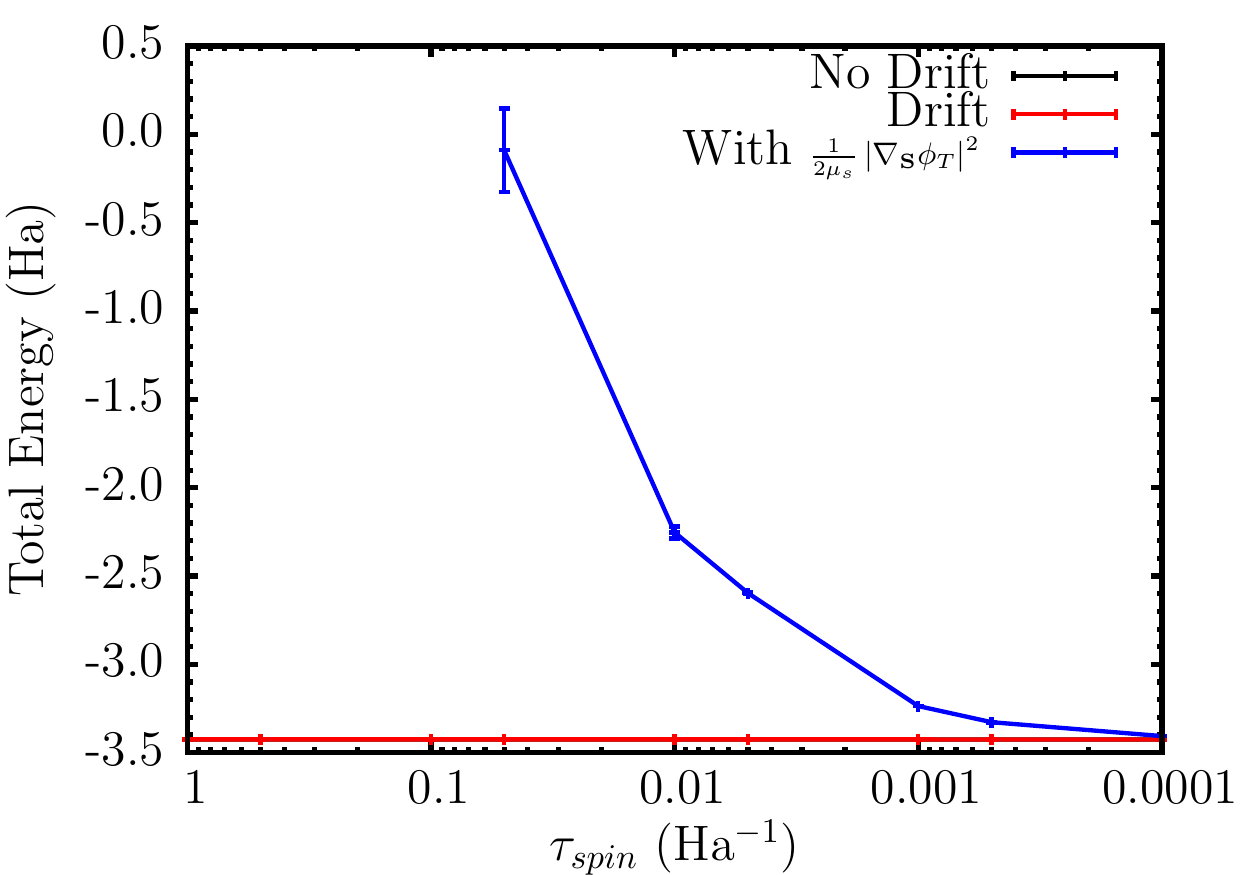}
\end{figure}

In order to determine the dependence on the spin mass/time-step, we studied the FPSODMC energy as a function of spin mass/time-step for a fixed spin time-step of the Pb atom ground state. In order to minimize the spatial time-step error, we choose a spatial time-step of $\tau = 0.001$~Ha$^{-1}$ throughout. Our trial wave functions are complete open-shell configuration interaction (COSCI), with one-particle spinors obtained from the DIRAC relativistic quantum chemistry code \cite{dirac14}. Additionally, we performed calculations that include and exclude the spin drift velocity.  This amounts to a choice of when the importance sampling transformation is invoked in the algorithm. With the inclusion of the spin drift, the spin variables are updated with 
\begin{equation}
    s_i' = s_i + \sqrt{\tau_s}\eta + \tau_s v_D^{s_i}
\end{equation}
where $\eta$ is a normally distributed random variable and the greens function in equation (\ref{eqn:spin_space_greens}) is used. Excluding the spin drift amounts to the condition $\mathbf{v}_D^\mathbf{S} = 0$. 

We first justify dropping the term $1/2\mu_s\left| \nabla_\mathbf{S} \phi_T(\mathbf{X}) \right|^2$ from the local energy. This term arises from the importance-sampling transformation when we include the spin kinetic energy, which was introduced to allow for efficient sampling of the spin degrees of freedom. Clearly, this term is spurious since spins do not have kinetic energy. In order to illustrate its effect, a plot of the total energy of Pb ground state is shown in Figure \ref{figure:spin_phase_grad}. As $\tau_s$ gets large, $\mu_s$ approaches zero and fully integrates the spin degrees of freedom. However, since the spin phase gradient is proportional to $\mu_s^{-1}$, it adds a positive term to the local energy with rapidly increasing value for larger spin time steps. From here on, all calculations are performed without this term. 

\begin{figure}[!t]
    \caption{Total energies of the Pb ground state with COSCI trial wave functions. The calculations were performed with drift, $\mathbf{v}_D^{\mathbf{S}} \ne 0$, and excluding drift, $\mathbf{v}_D^{\mathbf{S}}$ = 0.} 
    \centering
    \label{figure:Pb_drift_nodrift}
    \includegraphics[width=0.5\textwidth]{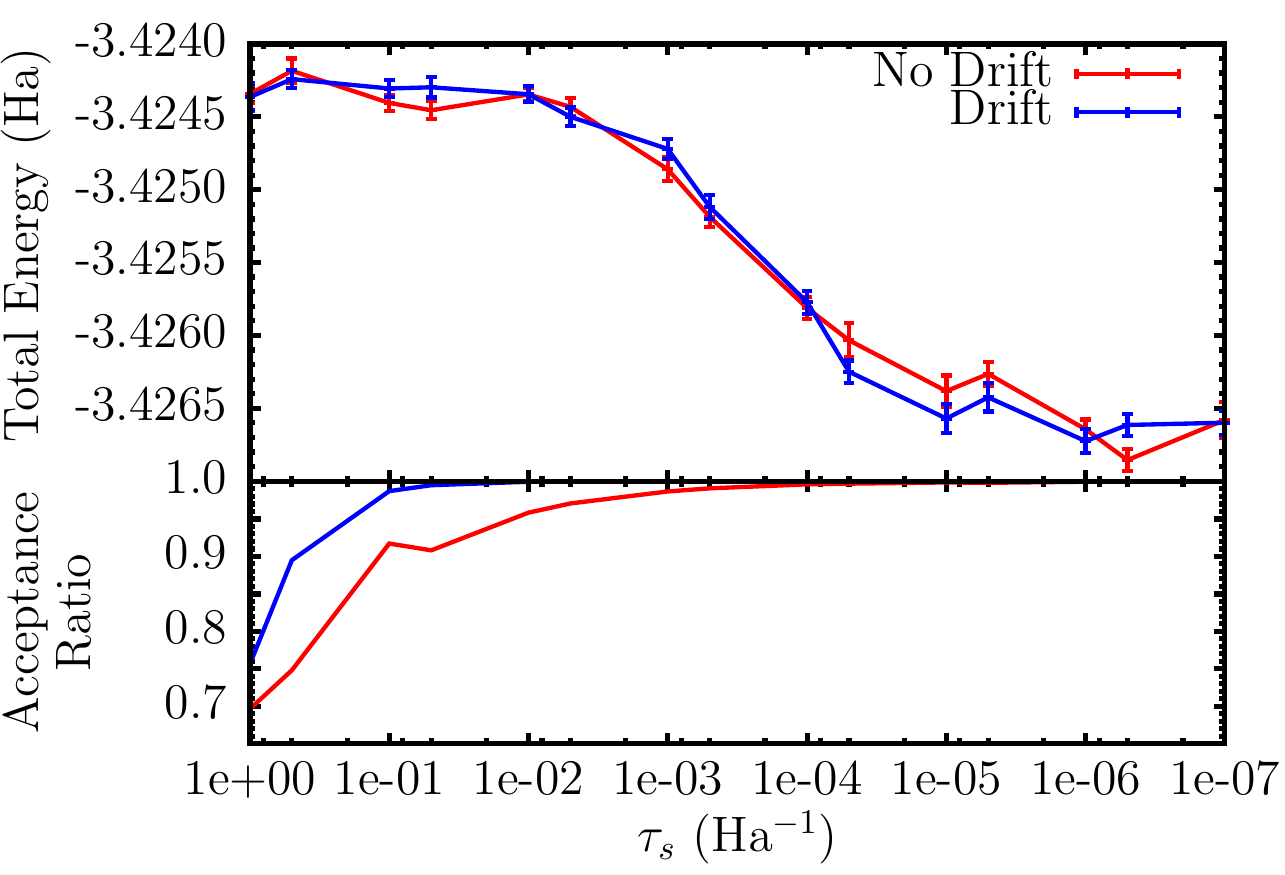}
\end{figure}
The dependence on the spin drift velocity is shown in Figure \ref{figure:Pb_drift_nodrift}. For large spin timesteps and small spin masses, the total energies for both methods saturate at $\sim -3.4244$~Ha. For spin timesteps $\tau_s \ge 1.0$~Ha$^{-1}$, the acceptance ratios drop below 0.7. In this regime, the method is no longer purely DMC. For spin timesteps $\tau_s < 1.0$~Ha$^{-1}$, the energies agree between the with/without spin drift calculations to within the error bars. The inclusion of the spin drift acts to increase the acceptance ratio; for spin timesteps larger than $\tau_s = 0.1$~Ha$^{-1}$, the acceptance ratio is greater than 0.99, which is desired for a DMC calculation. An interesting feature is a small decrease in total energy near $\tau_s = 0.001$~Ha$^{-1}$.
 For very small spin timesteps, the energy saturates at $\sim -3.4265$~Ha. 

\begin{figure}[!t]
    \caption{Total energy using COSCI and CISDT trial wave functions for the Pb ground state. FCI with cc-VQZ and a CBS extrapolation are included as a reference.}
    \label{figure:Pb_cosci_sdt}
    \centering
    \includegraphics[width=0.5\textwidth]{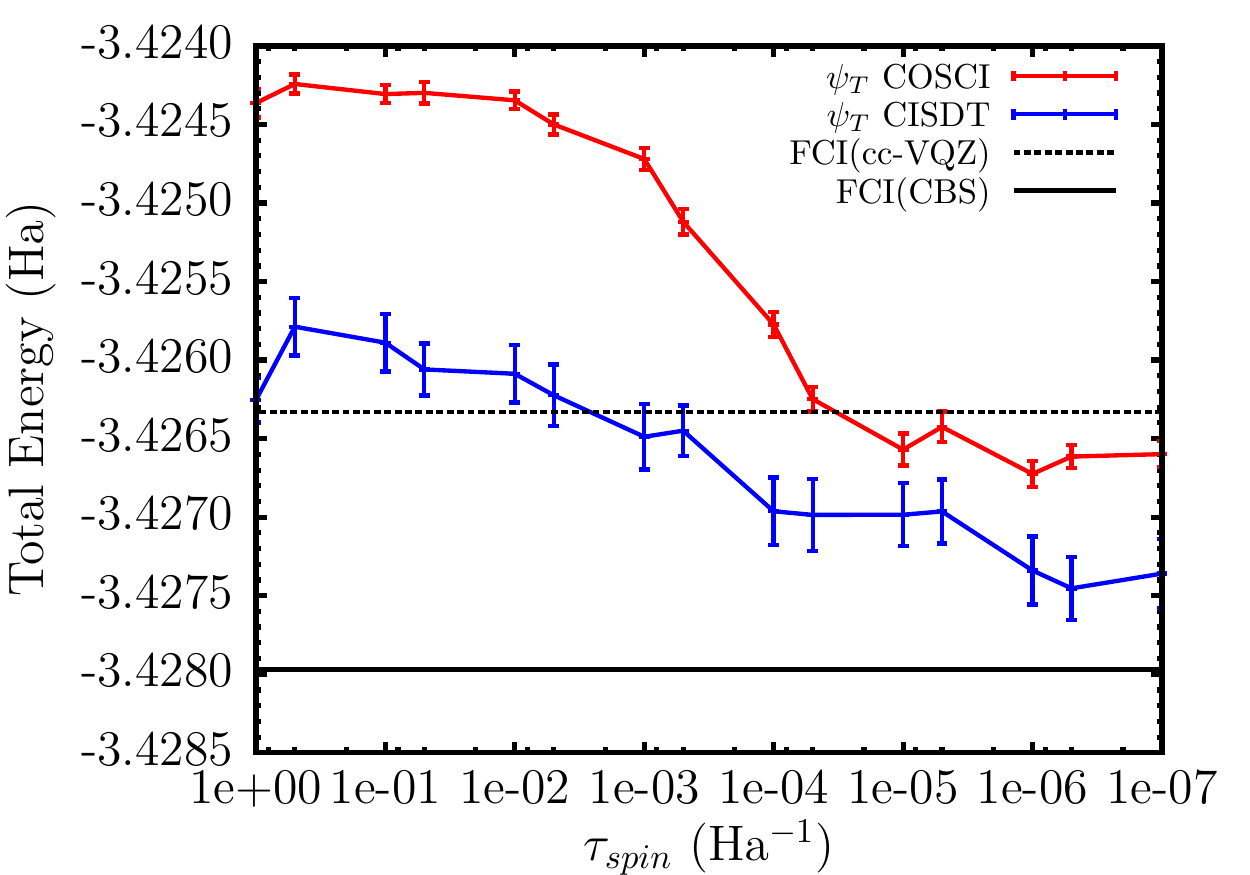}
\end{figure}

In order to determine how the energies compare to the exact eigenvalue for this Hamiltonian, we performed full configuration interaction (FCI) calculations with cc-V$n$Z basis sets and extrapolate to the complete basis set limit (CBS), using the same effective Hamiltonian within the two-component spinor formalism. For extrapolation, we use fits of the form 
\begin{eqnarray}
    E_{CBS} &=& E_{CBS}^{f(\textrm{COSCI})} + E_{CBS}^{g(\textrm{FCI-COSCI})} \\
    f(x) &=& E_{CBS}^{f(x)} + \alpha e^{-\beta n} \\
    g(x) &=& E_{CBS}^{g(x)} + \frac{\gamma}{(n-3/8)^3} +
    \frac{\delta}{(n-3/8)^5}
\end{eqnarray}
where $n$ refers to the size of the basis set and $\alpha$, $\beta$,
$\gamma$, and $\delta$ are fitting parameters. In addition to a COSCI trial wave function, we also calculated the total energy using a CISDT trial wave function for the ground state. The results are shown in Figure \ref{figure:Pb_cosci_sdt}. For $\tau_s < 0.0001$~Ha$^{-1}$, both $\psi_T$ used for the FPSODMC calculations are below the FCI calculations with a cc-VQZ basis set. The FCI with a CBS extrapolation is the best estimate for the exact ground state of this effective Hamiltonian, and the FPSODMC method lies above in both cases, due to the fixed-phase bias.

For a large spin time-step and small spin-mass, the spin configuration space is sampled faster than the spatial degrees of freedom. This corresponds to the plateau at -3.4244~Ha as seen in Figure \ref{figure:Pb_drift_nodrift}. However if $\tau_s$ gets too large relative to $\tau$, the spin steps become very large and lowers the acceptance ratio significantly. In the other regime, namely where $\tau_s \ll \tau$ and $\mu_s$ is large, the spin degrees of freedom diffuse slowly relative to the spatial degrees of freedom and are effectively slowly moving barriers. Therefore spatial degrees of freedom have time to find and favor the minimas subject to the fixed-phase given by the slowly moving spins. This drives the energy down (Figures \ref{figure:Pb_drift_nodrift} \& \ref{figure:Pb_cosci_sdt}), however it is still variationally bound with respect to the exact eigenvalue given from the FCI(CBS) energy. This is interesting to observe since due to overcomplete representation for the spins 
one cannot rule out that the algorithm in the limit of slow spin  evolution might find energies that would not be variational. However, all the indication are are that the fixed-phase approximation dominates over the full range of the spin timesteps. 


For the sake of completeness, we also performed FPSODMC calculations using the minimal spin representation, where the spin variables are randomly sampled from $s_i \in \left\{-1/2,1/2\right\}$. For the Pb atom with a LC REP, there are only $2^4 = 16$ spin configurations. For larger systems, the spin space grows as $2^N$. The spin variables were updated at each spatial step with uniform sampling, and we found an energy of -3.4239(3)~Ha that is comparable to results that we have obtained with our $S^1$ representation. 
However, the DMC acceptance ratio is only $\approx 0.92$  and clearly the acceptance can get significantly lower for larger systems (in fact, we expect exponential decrease of the acceptance in large systems). 
It appears to be challenging to counter this trend since discrete representation does not provide any drift that could boost the acceptance. Therefore our expanded spin representation is significantly more efficient and provides several other advantages as elaborated upon above. In addition, there might be further gains in improving our method that will be explored in subsequent work. 

\section{Applications}\label{section:applications}
We present applications of the FPSODMC method to several examples of electronic structure problems with significant impact of the spin-orbit terms. We apply the method to several electronic properties of atoms and molecules. In \S \ref{section:PbH}, we present results for the lead hydride PbH, where we present results for the binding energy and bond length. We perform both averaged scalar-relativistic and spin-orbit treatments of this molecule using FPSODMC to demonstrate the importance of spin-orbit on the electronic properties. In \S \ref{section:Sn}, we present a study of the Sn atom and dimer as an example of the 4th row element that shows significant impact of the spin-orbit. For the atom, we calculate the electronic structure of the first few excited states with remarkable agreement with experiment. For the dimer, we present scalar and spin-orbit treatments to predict the binding energy and bond length. Lastly in \S \ref{section:EA}, we calculate the electron affinities of the 6$p$ elements. All trial wave functions were constructed from one-particle spinors obtained from the DIRAC \cite{dirac14} code.

\subsection{PbH}\label{section:PbH}
We present calculations of  bond lengths and  dissociation energies of the  linear molecule PbH as a testing cases which can be compared with previous high-accuracy calculations. 
Previous theoretical studies have focused on two aspects of Pb  molecules: treatment of the correlation that requires multi-reference wave functions and relativistic effects including both scalar and beyond. Both of these are also addressed in our DMC calculations.

 \begin{table}
        \centering
        \normalsize
        \setlength{\tabcolsep}{3.5pt}
        \caption{PbH bond length ($r_e$) and dissociation energy ($D_e$) }
        \label{tbl:leadhydride}
  \renewcommand{\arraystretch}{1.1}
\begin{tabular}{ l  c  c }
\hline\hline
 Method     &$r_e$ (\AA)      &$D_e$ (eV)   \\[0.5ex] 
\hline 
~spin-free CCSD(T)$^{a}$ &1.836 & 2.66 $\;\;\;$  \\
~MRCIS-spss(CCSD(T))$^{b}$  &1.834 &1.61-1.71$\;\;$ \\ 
 ~$ \text{DMC/1-component AREP}$  &1.834 &2.582(3)\\
 ~$ \text{DMC/2-component REP}$ &1.838 &1.67(2)\\
 ~Exp\cite{exp_PbH/PbO} &1.839 &$\le1.69$ \qquad\qquad  \\[0.7ex]
\hline
\multicolumn{3}{l}{\small{~$^a$ 1-component CCSD(T) with large-core AREP}  \cite{STU2000} }\\[-1mm]
\multicolumn{3}{l}{\small{~$^b$ 2-component MRCIS with spin-free-state shift (spss) }}\\[-1mm]
\multicolumn{3}{l}{ \small{~~~~~$\,\,\,$evaluated at the 1-component CCSD(T) level}  \cite{STU2000} }
\end{tabular}
\end{table}

For PbH molecule, a large core (LC) REP\cite{STU2000} was found to be sufficiently accurate and the results 
are presented in Table \ref{tbl:leadhydride}. We contrast two types of DMC calculations with regard
to treatment of the spin degrees of freedom. 
One is the usual static-spin (1-component) calculation with weighted average of
the spin-orbit terms  for the same $l$ in the PP, denoted as AREP defined above. The second type
employs the present method with spinors and 
 genuine  2-component relativistic PP that includes SO interaction terms explicitly (REP).  
The result of REP calculations show excellent agreement with the experiment and previous SO-CI \cite{STU2000,exp_PbH/PbO} study. We point out the large discrepancy ($\approx$ 0.9 eV) 
 between the results of AREP and REP for the dissociation energy.
 It is clear that the averaged SO treatment is grossly inadequate for producing a reliable result for this system. The dominant correction for SO effects in this case comes from the atom, about  ~ 1 eV. The 2-component result is clearly more consistent with the experiment value. Note that the molecular spin-orbit correction $\rm \Delta_{SO}^{M}=E_{DMC}^{AREP}-E_{DMC}^{REP}$ 
for PbH molecule is estimated to be actually rather small, about 0.11 eV. However, in this case
the impact of the SO terms for Pb atom is dominant, about 1.1 eV. Thus, the atomic SO correction contributes most of the large discrepancy between the AREP and REP results.

\subsection{Sn and Sn$_2$}\label{section:Sn}
We investigate the effect of the spin-orbit interaction on Sn systems, namely the Sn atom and dimer. We investigate both large-\cite{STU2002} and small-core\cite{STU2000} PPs with 46 and 28 electrons removed respectively. In order to isolate the effect of spin-orbit, we perform static-spin calculations using only the PP via FNDMC as well as full dynamical spin calculations with the full REP via FPSODMC using COSCI trial wave functions. By comparing the AREP and REP calculations, we calculate an atomic spin-orbit correction $\Delta_\textrm{SO} = \textrm{E}_\textrm{DMC}^\textrm{AREP} - \textrm{E}_\textrm{DMC}^\textrm{REP} $ between the ground states. For the LC, we find a correction of $\Delta^\text{LC}_\textrm{SO} = 0.1689(2)$ eV and $\Delta^\text{SC}_\text{SO} = 0.27(2)$ eV for the SC. From Table \ref{table:Sn}, we see that the SC REP performs significantly better than the LC REP and agrees remarkably well with the experimental excitation energies. We also performed full configuration interaction (FCI) calculations using the LC REP in order to compare with our FPSODMC excitation energies. 

\begin{table}[!b]
\centering
\normalsize
\caption{Excitation energies for the Sn atom from the $^3P_0$ ground state. We include both LC and SC PPs. For completeness, we include COSCI and FCI to compare the FPSODMC and experiment\cite{NIST}.}
\label{table:Sn}
\begin{tabular}{c | c c |c c | c c| c}
\hline\hline
State & COSCI & DMC & COSCI & DMC & FCI$^\dagger$ & CISD$^\dagger$ &  Expt. \cite{NIST}\\
& LC & LC & SC & SC & LC & SC & \\
\hline
$^3P_1$ & 0.168 & 0.145(2) & 0.180 & 0.23(2) & 0.175 & 0.196 & 0.210\\
$^3P_2$ & 0.392 & 0.367(3) & 0.412 & 0.43(2) & 0.375 & 0.416 & 0.425\\
$^1D_2$ & 1.311 & 0.967(3) & 1.308 & 1.08(2) & 1.035 & 1.146 & 1.068\\
$^1S_0$ & 2.783 & 2.119(3) & 2.742 & 2.17(2) & 2.214 & 2.279 & 2.128\\
\hline
\multicolumn{5}{l}{ \small{$^\dagger$ cc-pVTZ } \cite{Sn-basis} }\\[-1mm]
\end{tabular}
\end{table}

For the dimer, we aim to find the equilibrium geometry as well as the binding energy. To determine the effect of spin-orbit on the dimer, we calculate the dimer both with the AREP and REP with COSCI trial wave functions. We perform calculations at various bond lengths and fit the binding curve to the Morse potential of the form 
\begin{equation}
    V(r) = D_e\left(e^{-2a(r-r_e)}-2e^{-a(r-r_e)}\right)
\end{equation}

\begin{figure}[!t]
    \caption{Binding Curve of the Sn$_2$ molecule using averaged spin-orbit AREP with FNDMC and  spin-orbit REP with FPSODMC methods. The curves are offset to dissociation limit $2E_0(\textrm{Sn})$ within each method to enable comparison for the predicted binding energy of each method with experiment. }
    \label{figure:Sn2}
    \centering
    \includegraphics[width=0.48\textwidth]{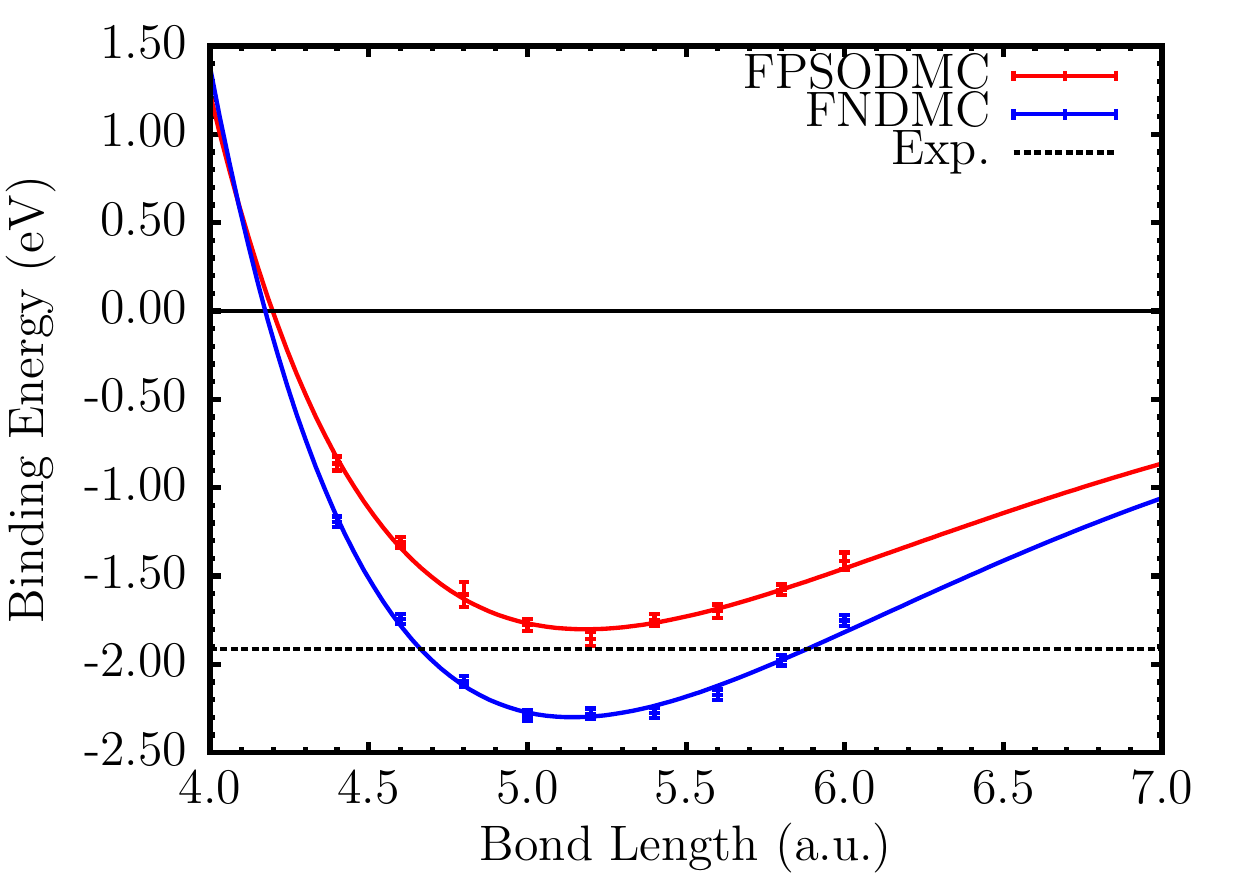}
\end{figure}

\begin{figure}[!b]
    \label{figure:Sn2_LC_SC}
    \caption{Binding curve of the Sn$_2$ molecule using large- and small-core REP. The large- and small-core systems have 8 and 44 valence electrons, respectively.}
    \centering
   \includegraphics[width=0.48\textwidth]{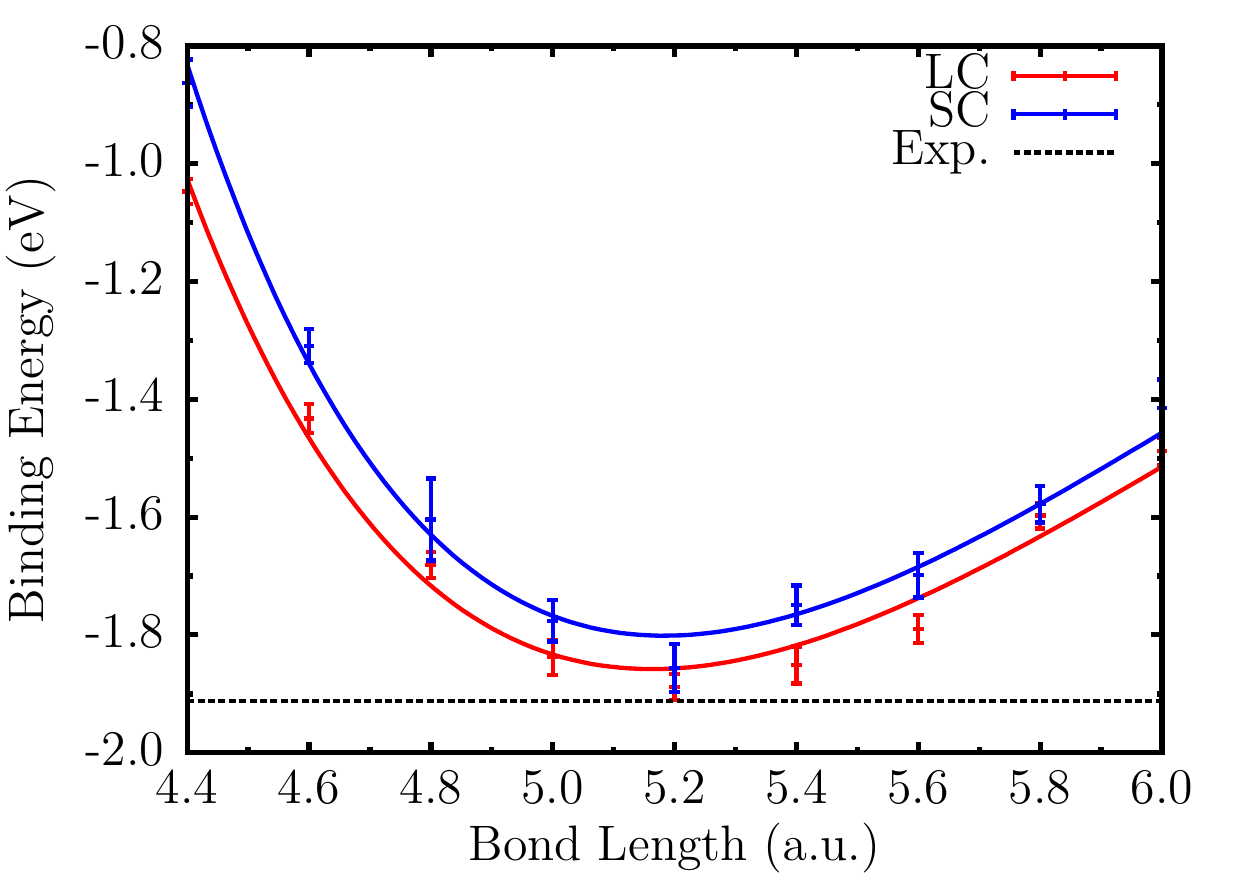}
\end{figure}

Results are shown in Figure \ref{figure:Sn2}. As was the case for PbH in \S \ref{section:PbH}, the scalar relativistic calculation without spin-orbit (AREP) significantly overbinds the dimer. The scalar relativistic method predicts an equilibrium bond length of $r_e = 2.74(2)$~\AA\, and a dissociation energy of $D_e = 2.32(3)$~eV. We note that these results are comparable to other DMC studies of the Sn$_2$ molecule with scalar relativistic pseudopotentials \cite{dmc-sn2}  For the dimer including the SO interaction, we find an equilibrium bond length of $r_e = 2.742(9)$~\AA\, and a dissociation energy of $D_e = 1.80(2)$~eV. We calculate a molecular spin-orbit correction $\Delta_{SO}^M $ = $E^{\textrm{AREP}}_{\textrm{DMC}} - E^{\textrm{REP}}_{\textrm{DMC}}$ at the equilibrium bond length of $\Delta_{SO}^M = 0.12(3)$~eV. We note that a COSCI-MVOQ scalar relativistic treatment predicts $r_e = 2.860$~\AA\, and a $D_e = 1.704 $~eV whereas COSCI with spin orbit treatment predicts a bond length of $r_e = 2.91$~\AA\, and a dissociation energy $D_e = 0.858$~eV. We also calculated the binding curve for the LC REP within FPSODMC with a COSCI trial wave function, shown in Figure \ref{figure:Sn2_LC_SC}. The LC REP produces an overall shift in the binding curve of $\sim$~0.1~eV. 

Experimental data for the dimer agrees remarkably well with the FPSODMC calculations. The experimental bond length was found to be $r_e^{\textrm{exp}} =2.748$~\AA\cite{Sn2expt} and a dissociation energy of $D^{\textrm{exp}}_e = 1.9125(31)$~eV \cite{Sn2expt}, which agree to our FPSODMC numbers to approximately 0.1~eV. We note that all DMC calculations (with and without SO) predict an equilibrium geometry that agrees with experiment to approximately 0.1~\AA. Due to the fact that Sn is from the fourth row, the spin-orbit correction to the atom and dimer energies are intermediate size (0.27(2)~eV and 0.12(3)~eV respectively), these contributions are not negligible if one desires predictions to chemical accuracy.  

\subsection{Electron Affinities}\label{section:EA}
As a last illustration of the method, we present results for the electron affinities (EA) of various atoms, namely the 6$p$ elements Tl-At. These elements have rather rather strong spin orbit coupling compared to their isovalent counterparts. For example, the respective EAs of C, Si, Ge and Sn are 1.2629~eV, 1.385~eV, 1.2~eV, and 1.2~eV\cite{hotop}. However, due to the strong spin-orbit effect on the $^3P_0$ ground state of Pb, the EA is significantly reduced which was experimentally determined to be 0.365(8)~eV \cite{eaPbExpt}. Theoretical calculations that did not include spin-orbit predict an EA of 1.284~eV, which is comparable to Ge and Sn where the spin-orbit interaction is significantly weaker\cite{peterson}. The inclusion of spin-orbit has a significant effect; a full four-component relativistic treatment predicts an EA of 0.403(39)~eV\cite{eaPbTheory} whereas FPSODMC using REPs predicts 0.417(7)~eV\cite{melton}. 

The electron affinity is calculated from the ground state energies of the neutral, $E_0$, and anionic $E_0^-$ species, namely
\begin{equation}
    EA = E_0 - E^-_0
\end{equation}
For each species, we calculate the EA using an AREP and REP to isolate the effect of spin-orbit. The results are presented in Table \ref{table:EA}. Electron affinity grows as proton number increases. The scalar relativistic (AREP) calculation for Bi does not follow this trend, and is significantly lower. For Pb, Po, and At, the EA is higher than experiment or other quantitative estimates. The inclusion of spin-orbit significantly improves these results to be accurate to within ~ 0.1 eV of the experimental values for Pb and Bi and in complete agreement with other quantitative estimates for Po and At where no experimental data exists to our knowledge.

\begin{table}[!t]
    \label{table:ea}
    \centering
    \caption{Electron Affinities for the 6$p$ elements. COSCI trial wave functions used throughout, with LC AREP/REPs for Pb\cite{STU2000}, Bi\cite{STU2002}, Po\cite{STU2002}, and At\cite{STU2002}. For Tl, no LC REP was found, so we utilize a SC AREP/REP \cite{TL-ECP}.}
    \label{table:EA}
    \begin{tabular}{c | c c|  c c | c }
    \hline\hline
    Species & COSCI & COSCI & FNDMC & FPSODMC & Expt\\
            & AREP & REP & AREP & REP \\
    \hline
    Tl & -0.015 & -0.195 & 0.29(2) & 0.17(3) & 0.377(13) \cite{eaTlExpt}\\
    Pb &  0.951 & -0.085 & 1.35(1) & 0.417(7)\cite{melton} & 0.365(8)\cite{eaPbExpt} \\
    Bi & -0.144 &  0.080 & 0.82(1) & 1.04(2) & 0.942362(13) \cite{eaBiExpt}\\
    Po &  0.981 &  0.556 & 1.94(1) & 1.32(6) & 1.32$^\dagger$\\
    At &  2.280 &  1.545 & 3.22(1) & 2.83(8) & 2.80(2)$^\dagger$ \\
    \hline
    \multicolumn{6}{l}{ \small{$^\dagger$ No experimental data. Quantitative estimates} \cite{EAest} } 
    \end{tabular}  
\end{table}

\section{Conclusions}\label{section:conclusions}

In this paper we elaborate in detail on the fixed-phase spin-orbit DMC (FPSODMC) method that we have introduced recently \cite{melton}. We provide detailed derivations
for several aspects of the method. One important point is the proof of upper bound property when dealing with complex nonlocal operators that enables us to epmploy techniques based on T-moves, ie, using combination of projection on the trial function and nonlocal sampling. The next point we demonstrate is more practical and has to do with the time step biases with regard to spin sampling together with the overcompleteness of the representation. We provide calculations of several systems that illustrate the capabilities of the method as well as show 
the impact of the spin-orbit interactions
on energy differences. The method opens new perspectives for 
many-body electronic structure calculations spinor formalism
that take into account variable nature of the spin degrees of 
freedom. 

{\em Acknowledgments.}
This research was supported by the U.S. Department of Energy (DOE), Office of Science, Basic Energy Sciences (BES) under Award de-sc0012314. For calculations we used resources of the National Energy Research Scientific Computing Center, a DOE Office of Science User Facility supported by the Office of Science of the U.S. Department of Energy under Contract No. DE-AC02-05CH11231
as well as additional allocation at ANL Mira machine. Part of the calculations have been carried out also at TACC.

\end{document}